\documentclass[11pt,twoside,letterpaper]{article} 

\usepackage{times,fancyhdr}

\usepackage[dvips]{graphicx}

\usepackage{epsfig}

\usepackage{amssymb}

\usepackage{amsmath}

\usepackage{amsfonts}

\usepackage{amsthm,amscd}

\usepackage{amsbsy}

\usepackage{latexsym}

\usepackage{bm}

\usepackage{url} 			

\usepackage{layout}

\usepackage{fleqn} 		

\usepackage[square,numbers,sort&compress]{natbib}


\makeatletter 

\def\cleardoublepage{\clearpage\if@twoside \ifodd\c@page\else%

    \hbox{}%

    \thispagestyle{empty}%

    \newpage%

    \if@twocolumn\hbox{}\newpage\fi\fi\fi} 

\makeatother

\def\figurename{Figure}

\makeatletter

\renewcommand{\fnum@figure}[1]{\figurename~\thefigure.}

\makeatother

\def\tablename{Table}

\makeatletter

\renewcommand{\fnum@table}[1]{\tablename~\thetable.}

\makeatother

\begin{document}

\title{ 
{\begin{flushleft}
\vskip 0.45in
{\normalsize\bfseries\textit{
	}}
\end{flushleft}
\vskip 0.45in
\bfseries\scshape Nonlinear Limits to Optomechanical Thermometry}
}

\author{\bfseries\itshape Sina Khorasani \thanks{E-mail address: sina.khorasani@ieee.org}\\
Vienna Center for Quantum Science and Technology, \\University of Vienna, Vienna, Austria}

\date{}

\maketitle

\begin{abstract}
	Optomechanical thermometry is a precise and reference-free method to measure absolute temperature. While pumping high optical power is needed to overcome noise and reduce the integration time, there is actually an upper limit to the useful optical power regardless of all other nonideal effects. Side-band inequivalence is a nonlinear effect obtained by higher-order operator algebra in quantum optomechanics and equivalent experiments, which causes asymmetric frequency shifts in side-bands and also an additional difference in their population. This chapter discusses previously unnoticed nonlinear effects arising from side-band inequivalence in optomechanical and Raman thermometry, which determines an upper bound in available optical power for temperature readout.
\end{abstract}

\vspace{.08in} \noindent \textbf{Keywords:} Operator algebra, Nonlinear interactions, Higher-order Operators, Side-band inequivalence, Quantum optomechanics, Raman spectroscopy.

\thispagestyle{empty}

\setcounter{page}{1}


\thispagestyle{fancy}

\fancyhead{}

\fancyhead[L]{In: Understanding the Schr\"{o}dinger Equation \\ 
Editor: Editor Name, pp. {\thepage-\pageref{lastpage-01}}} 
\fancyhead[R]{ISBN 0000000000  \\
\copyright~2020 Nova Science Publishers, Inc.}
\fancyfoot{}

\renewcommand{\headrulewidth}{0pt}


\vspace{2in}


\pagestyle{fancy}

\fancyhead{}

\fancyhead[EC]{Sina Khorasani}

\fancyhead[EL,OR]{\thepage}

\fancyhead[OC]{Nonlinear limits to Optomechanical thermometry}

\fancyfoot{}

\renewcommand\headrulewidth{0.5pt} 


\section{Introduction} \label{Section-1}

Optomechanics \cite{Kip,Aspel,Bowen} is a rapidly growing field in modern quantum optics, which has been constantly challenging the ultimate fabrication methods, and pushing the experimental boundaries and theoretical techniques forward in quantum noise to the extremes. Nonlinearity is an essence of optomechanical interactions, where radiation pressure of electromagnetic field with frequency $\omega$ acting upon a vibrating reflective mirror with frequency $\Omega$ creates a combination of nonlinear action of vibrating mirror upon light, and nonlinear backaction of optical field upon mechanical motion. As a result, the reflected scattered light from optomechanical cavity will contain frequency shifted components, which can be assumed to be placed approximately at mechanical frequency harmonics of the input electromagnetic light. Hence, the output spectrum will contain multiple peaks in pairs, which are placed \textit{almost} symmetrically around the central peak as $\omega\pm j \Omega$. The two closest ones to the center with $j=1$ form the first-order pair of side-bands with approximate frequencies $\omega\pm\Omega$. 

This particular kind of nonlinear interaction is not unique to optomechanics, and is also found in other equivalent experimental configurations with very different setups. These include superconducting electromechanics, ion traps, Paul traps, electrooptic modulators, acoustooptic modulators, Brillouin scattering, and Raman scattering. Among these, the last two must contain summations over phonon spectra, and could be a lot more complex in their most accurate description, however, for a given pair of side-bands normally approximation with one single phonon mode is quite sufficient. Ion and Paul traps operate at very low frequencies, typically on the order of few $100{\rm kHz}$, superconductive electromechanics work in the range of $1-10{\rm MHz}$, optomechanical cavities depending on the design could operate anywhere between $10{\rm MHz}$ up to a few $10{\rm GHz}$, Brillouin scattering is mostly noticeable at similar frequency range of $1-10{\rm GHz}$, electrooptic and acoustooptic modulators allow bandwidths up to a few $10{\rm GHz}$, and Raman scattering occurs noticeably at optical frequencies in the much higher frequency range of $1-10{\rm THz}$. Hence, various experimentally observable versions of optomechanics cover a rather wide range of electromagnetic spectrum, and as a matter of fact, they all share almost identical behavior.

Since the governing equations are in general nonlinear, it is common to study the behavior of optomechanical systems in the fully linearized approximation, where random fluctuations are treated as if they had to be infinitesimal. This is quite a good approximation for the case of many experimental observations. But it also turns out to be insufficient when nonlinear interactions are taken into account and properly analyzed. For instance, second- and higher-order side-bands with $j\geq2$ cannot be explained without nonlinearity, which can be well observed in experiments. Beyond the linear approximation, there are still nonlinear effects which remain a matter of speculation and can be easily mistaken with bistability (only for cavity optomechanics), dynamical instability, or other non-ideal effects if not appropriately taken into account. 

The most prominent example of nonlinear effects is side-band inequivalence \cite{SciRep}, which explains that every pair of side-bands are not exactly centered at $\omega\pm j \Omega$, and there is actually a slight frequency difference. Furthermore, without considering the quantum occupancy effects and different scattering rates, the amplitudes of red and blue side-bands are not equal and also a bit different. This difference under stead-state thermal equilibrium is explained by different occupation of the side-bands due to Bose-Einstein statistics and minuscule difference in scattering rates unto the side-bands. This asymmetry can be used to recover temperature rather accurately using a heterodyne measurement and without reference through measurement of optical spectrum \cite{Bowen,Purdy1,Purdy2,Safavi,Marin,Raman}. Both optomechanical and Raman setups can be employed to measure temperature this way, however, quantum optomechanics normally gives more precise results and clean measurements. 

The measurement accuracy of optical spectrum depends on a few factors, and that also includes the integration time. Having a large enough signal-to-noise ratio needs long measurements as well as high optical powers. In general, pumping more power gives cleaner and more visible side-bands, leading to better readouts. However, it turns out through nonlinear analysis that this is not the whole story. Once the nonlinear effects kick in, the accuracy of temperature readout becomes questionable and this is not due to optical losses which warm up the cavity, or other nonideal effects. It is because of side-band inequivalence.

Side-band inequivalence not only shifts both of the side-bands towards red, but also, it causes an extra overpopulation of the red side-band more than what is allowed by Bose-Einstein statistics. This nonlinearity is completely different in nature from quantum mechanical statistics, which supposedly determine the amplitudes or populations of side-bands. Furthermore, it is independent of temperature and when the side-band inequivalence gets large enough, the thermometry is no longer possible, since quantum mechanical effects shall be dominated by classical nonlinearity. Therefore, there exists an upper bound to the practically available optical pump power due to purely nonlinear effects, beyond which optomechanical or Raman thermometry without consideration of nonlinear effects is impossible. 

This is a truly remarkable and very counter-intuitive conclusion that in principle quantum mechanical phenomena could be influenced and even completely masked off by  nonlinearity.

\section{Side-band Inequivalence} \label{Section-2}

Referring to the lower-frequency side-band as Stokes or red close to $\omega-j \Omega$, and the higher-frequency side-band as anti-Stokes or blue close to $\omega+ j \Omega$, the accurate frequency shifts of side-bands from pump $\omega$ with proper redefinition of signs for frequency shifts are 
\begin{equation}
\label{eq1}
\Delta_r=\Delta+\Omega+\frac{1}{2}\delta,
\end{equation}
for the red side-band and 
\begin{equation}
\label{eq2}
\Delta_b=\Delta-\Omega+\frac{1}{2}\delta,
\end{equation}
for the blue side-band, where $\Delta=\omega_c-\omega$ is frequency detuning from cavity resonance $\omega_c$. Following this particular notation of red and blue detunings respectively in (\ref{eq1}) and (\ref{eq2}), the red detuning $\Delta_r$ is mostly positive while blue detuning $\Delta_b$ is mostly negative.

Additionally, $\delta$ is the side-band inequivalence \cite{SciRep,PhysRep} and represents a non-linear symmetry breaking. We furthermore may define normalized dimensionless side-band inequivalence as 
\begin{equation}
\bar{\delta}=\frac{\delta}{\Omega},
\end{equation}
which can be also explained in percentage as $\bar{\delta}(\%)=\bar{\delta}\times100\%$ more conveniently. Under normal operating conditions, which is the case for Raman and optomechanical thermometry experiments, side-band inequivalence is always positive and $\bar{\delta}>0$ holds. This implies that the frequency asymmetry resulting from side-band inequivalence occurs in such a way that both side-bands tend to move towards red. 

Side-band inequivalence would be much easier to resolve experimentally if \begin{equation}
\label{eq3}
\bar{\delta}>\frac{\beta}{\Omega},
\end{equation} 
holds, where $\beta$ is the measured linewidth of side-bands, related to $\Gamma$ and $\kappa$ respectively being the mechanical and optical decay rates. For resolved side-band cavities $\beta\approx\Gamma$. We refer to (\ref{eq3}) as the condition of visibility. Side-band inequivalence can be still observed if (\ref{eq3}) is not satisfied, but under cases where this condition is met, then side-band inequivalence is unmistakably large and quite visible. 

For side-band resolved cavities where $\kappa<\Omega$, it happens mostly that the approximation $\beta\approx\Gamma$ is a good one, since $\kappa>>\Gamma$. Eventually for the much less frequent and so-called case of reverse dissipation regime, $\Gamma$ could exceed $\kappa$. In case of Doppler cavities with $\kappa>\Omega$, either side-bands essentially do not form or are too small to be distinguishable from background noise, and therefore (\ref{eq3}) is irrelevant.

Interestingly, besides numerous experimental evidence presented in the preceding article \cite{SciRep} and its supplementary information, side-band inequivalence in frequency is clearly seen in the heterodyne thermometry measurement \cite{Purdy1}, where red and blue detuning frequencies are easily measureable to yield $\Delta_r=3.633{\rm GHz}\pm 10.4{\rm kHz}$ and $\Delta_b=-3.623{\rm GHz}\pm10.4{\rm kHz}$. These numbers are equivalent to a side-band inequivalence of $\delta=10{\rm MHz}\pm 20.8{\rm kHz}$ or $\bar{\delta}=0.27\%$. At the same time, the mechanical linewidth is $\Gamma=396{\rm kHz}$ close to the measureable linewidth of side-bands which is $\beta\approx0.410{\rm MHz}$, and hence the condition of visibility is also by far satisfied.

\subsection{Nonlinear Frequency Asymmetry} \label{Section-2-1}

A full nonlinear analysis of side-band inequivalence \cite{SciRep,PhysRep} has been carried out in a recent study, showing that $\bar{\delta}$ can be well approximated as
\begin{equation}
\label{eq4}
\bar{\delta}\approx\frac{2\Gamma^2+8g^2}{\gamma^2+4\Omega^2\left[1-2\left(\frac{g}{\Omega}\right)^2\right]^2},
\end{equation}
in which $\gamma=\kappa+\Gamma$ and $g$ is the enhanced optomechanical interaction rate, expressed as
\begin{equation}
g=g_0\sqrt{\bar{n}}.
\end{equation}
Here, $g_0$ is the single-photon optomechanical interaction rate and $\bar{n}$ is the intracavity photon number, determined by optical pump power. Relationship (\ref{eq4}) is a good approximation if value of $g/\Omega$ is not close to $1/\sqrt{2}$. For $g<<\Omega$ in weak coupling limit and a side-band resolved cavity (\ref{eq4}) behaves as 
\begin{equation}
\label{Weak}
\bar{\delta}\approx\frac{2g^2}{\Omega^2}=\frac{2g_0^2}{\Omega^2}\bar{n}.
\end{equation}
For $g>\Omega$ corresponding to the strong coupling limit and a side-band resolved cavity (\ref{eq4}) becomes
\begin{equation}
\label{Strong}
\bar{\delta}\approx\frac{\Omega^2}{2g^2}=\frac{\Omega^2}{2g_0^2}\frac{1}{\bar{n}}.
\end{equation}

When the system is sufficiently away from optical bistability, one may use the approximation 
\begin{equation}
\label{eq7}
\alpha\approx\sqrt{\frac{\eta \kappa P_{\rm op}}{\hbar\omega}},
\end{equation} 
for the incident photon flux $\alpha$ with $\eta$ being the external coupling efficiency normally on the order of $0.1$ and $P_{\rm op}$ is the optical pump power. Knowledge of the incident photon flux $\alpha$ will determine the intracavity photon number $\bar{n}$ through solution of the third-degree equation
\begin{equation}
\label{eq8}
\alpha^2=\bar{n}\left[\frac{\kappa^2}{4}+\left(\frac{2g_0^2\Omega}{\Omega^2+\frac{1}{4}\Gamma^2}\bar{n}+\Delta\right)^2\right],
\end{equation}
which has exactly one real positive root for $\bar{n}$ if $\Delta\geq0$. It is not difficult to identify a critical blue detuning $\Delta_B<0$ through solution of another third-degree polynomial equation as the onset of bistability for all $\Delta<\Delta_B$. 

Hence, it is to be noticed that $\bar{\delta}$ in (\ref{eq4}) is actually a strongly varying function of optical power, with which it exhibits a resonant behavior when $g=\Omega/\sqrt{2}$, or $\bar{n}=\Omega^2/2g_0^2$. This condition is not actually quite straightforward to satisfy since bistability also starts to show up near the same pumping level, unless pump is red-detuned with $\Delta\geq0$ which is the sufficient, and not necessary, condition for its absence.

For the moment being, let us assume that mechanical quality factor is large $\Omega>>\Gamma$ and move forward with the case of resonant pump $\Delta=0$, which furthermore eliminate the optical spring effect, too. This will transform (\ref{eq7}) and (\ref{eq8}) into 
\begin{equation}
\label{eq9}
\bar{n}\left(\frac{\kappa^2}{4}+\frac{4g_0^4}{\Omega^2}\bar{n}^2\right)=\frac{\eta \kappa P_{\rm op}}{\hbar\omega}.
\end{equation}
There will be two limiting cases for weak and strong pump with the solutions obtained after a bit of effort as
\begin{equation}
\label{eq10}
\bar{n}\approx\left\{\begin{matrix}
\frac{4\eta }{\hbar\omega\kappa}P_{\rm op},&P_{\rm op}<<\mathcal{A}^2\kappa\hbar\omega/\eta,\\
\sqrt[3]{\frac{\mathcal{A}\eta \Omega  }{\hbar\omega g_0^2}}\sqrt[3]{P_{\rm op}},&P_{\rm op}>>\mathcal{A}^4\kappa\hbar\omega/\eta,
\end{matrix}
\right.
\end{equation}
where $\mathcal{A}=\Omega\kappa/4g_0^2$ needs to obviously satisfy $\mathcal{A}>1$ for (\ref{eq10}) to make sense. Now, plugging in (\ref{eq10}) in (\ref{Weak}) and (\ref{Strong}) respectively gives
\begin{equation}
\label{eq13}
\bar{\delta}\approx\frac{8g_0^2\eta }{\hbar\omega\kappa\Omega^2}P_{\rm op},
\end{equation}
for the weak coupling limit and 
\begin{equation}
\label{eq14}
\bar{\delta}\approx\sqrt[3]{\frac{\hbar\omega\Omega^4}{2\eta g_0^2 \kappa}}\frac{1}{\sqrt[3]{P_{\rm op}}},
\end{equation}
for the strong coupling limit. Therefore, the side-band inequivalence initially increases proportionally to optical power as $\bar{\delta}\propto P_{\rm op}$ before starting to fade out at much higher optical power levels as $\bar{\delta}\propto 1/\sqrt[3]{P_{\rm op}}$.

\subsection{Nonlinear Amplitude Asymmetry} \label{Section-2-2}

The nonlinear asymmetry of side-bands, or side-band inequivalence, does not end up with only the frequency asymmetry. It has been shown that side-band inequivalence causes additional asymmetry in the population of side-bands as well. This is very surprising result, since side-band inequivalence actually has two different and related behavior. Both asymmetries in frequencies and amplitudes lean towards red side-band. 

Extensive calculations using higher-order operators \cite{PhysRep} lead to the compact result \cite{SciRep}
\begin{equation}
\label{eq15}
\bar{n}_{r}-\bar{n}_{b}\approx \bar{n}\bar{\delta}.
\end{equation}
Here, $\bar{n}_{r}$ and $\bar{n}_{b}$ respectively correspond to the population of photons in the first red and blue side-bands. This quantity is directly measureable by recording the spectral noise density at each of the side-bands, while assigning $\frac{1}{2}$ to the shot-noise level. Precision of this measurement can be improved by prolonged observation of side-bands through multiple trace records of the spectral density and making an average in the end.

Since we always have $\bar{\delta}>0$, then we always can expect $\bar{n}_{r}>\bar{n}_{b}$. In the absence of frequency side-band inequivalence,  (\ref{eq15}) demands $\bar{n}_{r}=\bar{n}_{b}$ unless we take the quantum effects of Bose-Einstein statistics into account as well. This will be explained in the next section.

For the moment being, let us examine and investigate the limiting cases of (\ref{eq15}) under weak and strong coupling limits. For the weak coupling limit from (\ref{eq10}) and (\ref{eq13}) we get
\begin{equation}
\label{eq16}
\bar{n}_{r}-\bar{n}_{b}\approx\left(\frac{2\eta^2 }{\hbar^2\omega^2 \mathcal{A}^2\kappa\Omega}\right)P_{\rm op}^2.
\end{equation}
This explains that in the weak coupling limit, there exists an amplitude asymmetry between red and blue side-bands which must increase quadratically with optical power as
\begin{equation}
\label{eq17}
\Delta\bar{n}\propto P_{\rm op}^2.
\end{equation}
Here, we have defined $\Delta\bar{n}=\bar{n}_r-\bar{n}_b$. This is a rather important result. Later we shall observe it being in strong contrast with the results of quantum mechanical distributions. 

Meanwhile, the strong coupling limit using (\ref{Strong}) and (\ref{eq14}) gives
\begin{equation}
\label{eq18}
\Delta\bar{n}\approx\frac{\Omega^2}{2g_0^2}.
\end{equation}
This value shall set an upper bound to the maximum expected amplitude asymmetry because of nonlinear side-band inequivalence.

\section{Optomechanical Thermometry} \label{Section-3}

Under thermal equilibrium, the illumination of an optomechanical cavity with pumping light generates two side-bands. We proceed only with the first-order side-bands and we may note that side-band inequivalence in frequency does not change the frequency separation of side-bands because of (\ref{eq1}) and (\ref{eq2}). Hence, the populations of red and blue side-bands at a finite absolute temperature $T$ then must obey \cite{Bowen,Purdy1}
\begin{equation}
\label{eq19}
\frac{\bar{n}_b}{\bar{n}_r}=\exp\left(-\frac{\hbar \Omega}{k_{\rm B}T}\right),
\end{equation}
simply because of Bose-Einstein statistics, where $k_{\rm B}$ is Boltzmann's constant. This equation is the basic relationship to optomechanical thermometry where the populations of side-bands are measured first, and then by taking the logarithm of their ratio one may recover the absolute temperature $T$. Obviously, the optomechanical cavity must be placed in thermal contact and heat exchange with the sample to be observed. Additionally, thermal expansion and contraction contribute to small shifts in the mechanical frequency $\Omega$, cavity resonance $\omega_c$, and coupling ratio $\eta$, so that these effects also contribute to nonideal behavior which are already well known. 

\subsection{Quantum Amplitude Asymmetry} \label{Section-3-1}

At a fixed temperature where all non-ideal behavior can be ignored, we are only left with the optical power $P_{\rm op}$ to play with. In the weak coupling regime where linear approximation crudely applies, $\bar{n}_b\propto \bar{n}$ and $\bar{n}_r\propto \bar{n}$ hold to a high accuracy, so that we may rewrite  (\ref{eq19}) as \cite{SciRep}
\begin{equation}
\label{eq20}
\bar{n}_r-\bar{n}_b\propto\bar{n}.
\end{equation}
Using (\ref{eq10}) we may note that this results in 
\begin{equation}
\label{eq21}
\Delta\bar{n}\propto P_{\rm op}.
\end{equation}
This implies that at very low optical powers, the amplitude asymmetry will be linearly proportional to $P_{\rm op}$. This is the hallmark of a quantum effect, which dominates the side-band asymmetry and allows correct recovery of temperature. By increasing optical power, however, the side-band asymmetry shall be dominated by side-band inequiavalence in amplitude, which leads to a quadratic proportionality to $P_{\rm op}^2$. Once this regime is reached, quantum thermometry will be out of question. By continuing to increase the optical power, and in absence of other non-ideal effects, one may expect that the asymmetry be saturated at the constant level given by (\ref{eq18}).

\section{Nonlinear Limits} \label{Section-4}

It is possible to obtain the cross-over value for optical power $P_{\rm op}$ at which the transition from quantum to classical behavior takes place. This requires knowledge of proportionality constants in (\ref{eq20}). This is given first by taking note of the spectra density of a heterodyne measurement as \cite{Bowen}
\begin{equation}
\label{eq22}
S_{\rm het}(w)=\frac{1}{2}+S_{RR}(w)+S_{BB}(w),
\end{equation}
with $S_{RR}(w)$ and $S_{BB}(w)$ respectively being the contributions of side-bands on the red and blue sides to the spectrum, and $\frac{1}{2}$ is the background shot noise level. These are given as \cite{Bowen}
\begin{eqnarray}
\label{eq23}
S_{RR}(w)&=&\eta\Gamma |C_{\rm eff}(w+\Delta)|S_{QQ}(w+\Delta),\\ \nonumber
S_{BB}(w)&=&\eta\Gamma |C_{\rm eff}(w-\Delta)|S_{QQ}(w-\Delta),
\end{eqnarray}
with 
\begin{eqnarray}
\label{eq24}
S_{QQ}(w>0)&=&2\Gamma|\chi(w)|^2\left[m_{\rm th}+|\mathcal{C}_{\rm eff}(w)|+1\right], \\ \nonumber
S_{QQ}(w<0)&=&2\Gamma|\chi(w)|^2\left[m_{\rm th}+|\mathcal{C}_{\rm eff}(w)|\right],
\end{eqnarray}
being the mechanical spectral density. Here, the mechanical response function denoted by $\chi(w)$ is 
\begin{equation}
\label{eq25}
\chi(w)=\frac{\Omega}{\Omega^2-w^2-i w \Gamma}.
\end{equation}
Also, the thermal phonon occupation number is given by
\begin{equation}
\label{eq26}
m_{\rm th}=\frac{1}{\exp\left(\frac{\hbar \Omega}{k_{\rm B}T}\right)-1}.
\end{equation}
Furthermore, we have
\begin{equation}
\label{eq27}
\mathcal{C}_{\rm eff}(w)=\frac{\mathcal{C}}{\left(1-2i\frac{w}{\kappa}\right)^2},
\end{equation}
in which $\mathcal{C}=\bar{n}\mathcal{C}_0$ is the enhanced cooperativity and
\begin{equation}
\label{eq28}
\mathcal{C}_0=\frac{4g_0^2}{\kappa\Gamma},
\end{equation}
represents the single-photon cooperativity.

The photon population in each of the side-bands can be found by evaluating the spectral densities at their corresponding resonances. Hence, we have
\begin{eqnarray}
\label{eq29}
\bar{n}_r&=&S_{RR}(+\Omega), \\ \nonumber
\bar{n}_b&=&S_{BB}(-\Omega),
\end{eqnarray}
before considering the effect of side-band inequivalence in frequency. At resonant drive with $\Delta=0$, we obtain from (\ref{eq23}), (\ref{eq24}), (\ref{eq25}), and (\ref{eq27}) the appropriate expressions for the spectra of side-bands $S_{RR}(w)$ and $S_{BB}(w)$. These are
\begin{eqnarray}
\label{eq30}
S_{RR}(w)&=& \frac{2\mathcal{C}\eta\Gamma^2\kappa^2\Omega^2}{(\kappa^2+4w^2)^2}\frac{(m_{\rm th}+1)(\kappa^2+4w^2)+\mathcal{C}\kappa^2}{(\Omega^2-w^2)^2+ w^2 \Gamma^2},\\ \nonumber
S_{BB}(w)&=& \frac{2\mathcal{C}\eta\Gamma^2\kappa^2\Omega^2}{(\kappa^2+4w^2)^2}\frac{m_{\rm th}(\kappa^2+4w^2)+\mathcal{C}\kappa^2}{(\Omega^2-w^2)^2+ w^2 \Gamma^2}.
\end{eqnarray}
Hence, we get
\begin{eqnarray}
\label{eq31}
S_{RR}(+\Omega)&=&2\mathcal{C}\eta\kappa^2\frac{(m_{\rm th}+1)(\kappa^2+4\Omega^2)+\mathcal{C}\kappa^2}{(\kappa^2+4\Omega^2)^2},\\ \nonumber
S_{BB}(-\Omega)&=&2\mathcal{C}\eta\kappa^2\frac{m_{\rm th}(\kappa^2+4\Omega^2)+\mathcal{C}\kappa^2}{(\kappa^2+4\Omega^2)^2}.
\end{eqnarray}
These equations can be corrected to take account for the side-band inequivalence in frequencies on the red and blue side. Doing this will result
\begin{eqnarray}
\label{eq31a}
S_{RR}(+\Omega+\tfrac{1}{2}\delta)&\approx&S_{RR}(\Omega)+\frac{1}{2}\delta\frac{\partial}{\partial \Omega}S_{RR}(\Omega),\\ \nonumber
S_{BB}(-\Omega+\tfrac{1}{2}\delta)&\approx&S_{BB}(-\Omega)+\frac{1}{2}\delta\frac{\partial}{\partial (-\Omega)}S_{BB}(-\Omega),
\end{eqnarray}
from which we may obtain the normalized amplitude difference
\begin{eqnarray}
\label{eq32}
\Delta\bar{n}&=&\bar{n}_r-\bar{n}_b\\ \nonumber
&\approx&\frac{2\mathcal{C}_0\eta\kappa^2}{\kappa^2+4\Omega^2}\bar{n}+\frac{1}{2}\delta\frac{\partial}{\partial \Omega}\left[S_{RR}(\Omega)-S_{BB}(\Omega)\right] \\ \nonumber
&=&\frac{2\mathcal{C}_0\eta\kappa^2}{\kappa^2+4\Omega^2}\bar{n}-\frac{16\mathcal{C}_0\eta\kappa^2 g_0^2}{(\kappa^2+4\Omega^2)^2}\bar{n}^2\\ \nonumber
&=&\frac{8\mathcal{C}_0\eta^2\kappa}{\hbar\omega(\kappa^2+4\Omega^2)}P_{\rm op}-\frac{256\mathcal{C}_0\eta^3 g_0^2}{\hbar^2\omega^2(\kappa^2+4\Omega^2)^2}P^2_{\rm op}.
\end{eqnarray}
Here, by normalization of measured amplitudes assignment of $\frac{1}{2}$ to the background shot noise is implied. This equation contains two terms, first of which exhibits a linear dependence in amplitude asymmetry on optical power in agreement with (\ref{eq21}) due to mere quantum effects, as opposed to the quadratic one (\ref{eq17}) for classical nonlinearity coming from side-band inequivalence in amplitude. The second term arising from side-band inequivalence in frequency is however quadratic in optical power, and causes an apparent decrease in $\Delta\bar{n}$ at high optical powers.

\subsection{Quantum and Classical Asymmetries}

It is possible to merge (\ref{eq16}) and (\ref{eq32}) to obtain a better estimate to the normalized amplitude difference, giving rise to the polynomial expression
\begin{eqnarray}
\label{eq33}
\Delta\bar{n}&\approx&\left[\frac{8\mathcal{C}_0\eta^2\kappa}{\hbar\omega(\kappa^2+4\Omega^2)}\right]P_{\rm op}\\ \nonumber
&-&\frac{2\eta^2 }{\hbar^2\omega^2\mathcal{A}^2\kappa\Omega}\left[\frac{128\mathcal{C}_0\mathcal{A}^2\kappa\Omega\eta g_0^2}{(\kappa^2+4\Omega^2)^2}-1\right]P_{\rm op}^2.
\end{eqnarray}
This will mark a cross-over critical optical power at which transition from quantum to classical nonlinearity takes place. It is given by
\begin{eqnarray}
\label{eq34}
P_{\rm cr}&=&\frac{4\hbar\omega\mathcal{C}_0\kappa^2\Omega  \mathcal{A}^2}{\kappa^2+4\Omega^2}\left[\frac{128\mathcal{C}_0\mathcal{A}^2\kappa\Omega\eta g_0^2}{(\kappa^2+4\Omega^2)^2}-1\right]^{-1} \\ \nonumber
&=&\frac{\hbar\omega\kappa^3\Omega^3}{g_0^2\Gamma(\kappa^2+4\Omega^2)}\left[\frac{32\kappa^2\Omega^3\eta}{\Gamma(\kappa^2+4\Omega^2)^2}-1\right]^{-1} \\ \nonumber
&\approx&\frac{\hbar\omega\kappa(\kappa^2+4\Omega^2)}{32\eta g_0^2}.
\end{eqnarray}
For optical powers exceeding this limit with $P_{\rm op}>P_{\rm cr}$, classical nonlinearity dominates the quantum effect. For optical powers at lower levels with $P_{\rm op}<P_{\rm cr}$, quantum phenomena are still in effect. There are two terms within the brackets contributing to the critical optical power with opposite signs, the first of which comes from side-band inequivalence in frequency and the second of which comes from side-band inequivalence in amplitude. Here, the contribution of the former is dominant for most electromechanical and optomechanical setups, and causes decrease of $\Delta\bar{n}$  at $P_{\rm op}>P_{\rm cr}$. For thermometry applications, this in overall will cause underestimating temperature.

In equivalent terms, the critical intracavity photon number corresponding to the cross-over shall be given by
\begin{eqnarray}
\label{eq35}
\bar{n}_{\rm cr}&=&\frac{\kappa^2+4\Omega^2}{8 g_0^2}.
\end{eqnarray}
Finally, (\ref{eq34}) and (\ref{eq35}) in the limit of side-band resolved cavity $\Omega>\kappa$ can be approximated as
\begin{eqnarray}
\label{eq36}
P_{\rm cr}&\approx&\frac{\hbar\omega\kappa\Omega^2}{8\eta g_0^2}.
\end{eqnarray}
Similarly, we have
\begin{eqnarray}
\label{eq37}
\bar{n}_{\rm cr}&\approx&\frac{\Omega^2}{2 g_0^2}.
\end{eqnarray}

Taking the numbers for instance from \cite{Toth} with $\Omega=2\pi\times 5.33{\rm MHz}$, $\kappa=2\pi\times 118{\rm kHz}$, $\Gamma=2\pi\times 30{\rm Hz}$, $g_0=2\pi\times 60{\rm Hz}$, $\omega_c=2\pi\times 4.26{\rm GHz}$ and $\eta=0.76$.  This will give rise to a critical intracavity photon number $\bar{n}_{\rm cr}=3.94\times 10^{9}$. This corresponds to the incident optical power of only $P_{\rm cr}=0.1\mu{\rm W}$, which is well accessible experimentally. Even much larger values are possible in typical superconducting electromechanic setups, where an excessively large number of intracavity photons can be crunched into the cavity. 

Just as a cross-check, referring to (\ref{eq10}), the power threshold below which $\bar{n}\propto P_{\rm op}$ holds is too large $206{\rm mW}$, far above the values used in any optomechanical or electromechanical experiment. Henceforth, the proportionality $\bar{n}\propto P_{\rm op}$ holds with high precision at resonant pump. Similar calculations for typical optomechanical cavities using photonic crystal nanobeams \cite{Purdy1,Itay} gives exceedingly large values which are not practically accessible.

One should here notice that (\ref{eq37}) no longer satisfies the condition $g<<\Omega$ required in (\ref{Weak}), and this has to be taken care of in the original equation for side-band inequivalence given in (\ref{eq4}). Unfortunately, since (\ref{eq37}) implies resonant behavior in (\ref{eq4}) with $g/\Omega\approx 1/\sqrt{2}$, then (\ref{eq4}) is not a good approximation. However, one may expect some further enhancement in the results and change in the practical constraints.

Therefore, one alternative way to verify the effects arising from higher-order nonlinearities is to pump an electromechanical cavity on resonance with various optical powers, and observe the normalized amplitude difference on the red and blue side-bands. This difference should increase linearly with optical power up to a critical value, beyond which the linear increase will no longer hold. It is furthermore, advantageous to measure the cross-correlation function \cite{Purdy1} instead of the normalized amplitude difference

\section{Conclusions} \label{Section-5}

We presented an overview of side-band inequivalence, which causes two asymmetries in frequency and amplitude of side-bands. Both of these asymmetries happen to lean towards red, and as nonlinear effects depend on input pump power. While it seems that side-band inequivalence in frequency may have some effect on optomechanical thermometry using a heterodyne measurement, side-band inequivalence in amplitude causes a shift from linear to quadratic dependence, which ultimately causes classical nonlinearity of side-band inequivalence in amplitude to mask out and predominate the quantum asymmetry. These may set a constraint on the useful optical power which can be pumped into the cavity for precise thermometry. Furthermore, there appears to be a critical power above which quantum asymmetry is dominated by classical nonlinearity.

\section*{Acknowledgments}

This work is dedicated to the inspiring artist, Anastasia Huppmann.

\label{lastpage-01}

\end{document}